# The initial abundance and distribution of $^{92}$Nb in the Solar System

Tsuyoshi Iizuka [1]*, Yi-Jen Lai [2], Waheed Akram [2,3], Yuri Amelin [4], Maria Schönbächler [2,3]

[1] Department of Earth and Planetary Science, The University of Tokyo, Hongo 7-3-1, Bunkyo, Tokyo 113-0033, Japan

[2] Institute of Geochemistry and Petrology, ETH Zürich, Clausiusstrasse 25, 8092 Zürich, Switzerland

[3] School of Earth, Atmospheric and Environmental Sciences, The University of Manchester, Oxford Road, Manchester M13 9PL, UK

[4] Research School of Earth Sciences, The Australian National University, Canberra, ACT 2601, Australia

*To whom correspondence should be addressed.
E-mail: iizuka@eps.s.u-tokyo.ac.jp
Fax: +81 3 5841 8791
Tel: +81 3 5841 4282

**Abstract**

Niobium-92 is an extinct proton-rich nuclide, which decays to $^{92}$Zr with a half-life of 37 Ma. This radionuclide potentially offers a unique opportunity to determine the timescales of early Solar System processes and the site(s) of nucleosynthesis for *p*-nuclei, once its initial abundance and distribution in the Solar System are well established. Here we present internal Nb–Zr isochrons for three basaltic achondrites with known U-Pb ages: the angrite NWA 4590, the eucrite Agoult, and the ungrouped achondrite Ibitira. Our results show that the relative Nb–Zr isochron ages of the three meteorites are consistent with the time intervals obtained from the Pb–Pb chronometer for pyroxene and plagioclase, indicating that $^{92}$Nb was homogeneously distributed among their source regions. The Nb–Zr and Pb–Pb data for NWA 4590 yield the most reliable and precise reference point for anchoring the Nb–Zr chronometer to the absolute timescale: an initial $^{92}$Nb/$^{93}$Nb ratio of $(1.4 \pm 0.5) \times 10^{-5}$ at $4557.93 \pm 0.36$ Ma, which corresponds to a $^{92}$Nb/$^{93}$Nb ratio of $(1.7 \pm 0.6) \times 10^{-5}$ at the time of the Solar System formation. On the basis of this new initial ratio, we demonstrate the capability of the Nb–Zr chronometer to date early Solar System objects including troilite and rutile, such as iron and stony-iron meteorites. Furthermore, we estimate a nucleosynthetic production ratio of $^{92}$Nb to the *p*-nucleus $^{92}$Mo between 0.0015 and 0.035. This production ratio, together with the solar abundances of other *p*-nuclei with similar masses, can be best explained if these light *p*-nuclei were primarily synthesized by photodisintegration reactions in Type Ia supernovae.

# 1. Introduction

The proton-rich radionuclide $^{92}$Nb decays to $^{92}$Zr by electron capture with a half-life of 37 Ma (Holden, 1990). Since Nb and Zr can fractionate from each other during partial melting of the mantle, mineral crystallization and metal-silicate separation (Tiepolo et al., 2001; Wade and Wood, 2001; Klemme et al., 2002), the Nb–Zr system can potentially be used to determine the timescales of silicate differentiation and core segregation for infant planets (Minster and Allègre, 1982). In addition, the initial $^{92}$Nb abundance in the Solar System provides constraints on the nucleosynthetic site(s) of *p*-nuclei (*p*- denotes proton-rich) (Harper, 1996; Yin et al., 2000; Dauphas et al., 2003; Meyer, 2003; Hayakawa et al., 2013; Travaglio et al., 2014). These applications require the initial abundance and distribution of $^{92}$Nb (expressed as $^{92}$Nb/$^{93}$Nb) in the Solar System to be defined.

Evidence for live $^{92}$Nb in the early Solar System was first obtained from the iron meteorite Toluca in which rutile with high Nb/Zr exhibits $^{92}$Zr excesses (Harper, 1996). Assuming that the iron meteorite started with initial $^{92}$Nb/$^{93}$Nb and $^{92}$Zr/$^{90}$Zr values identical to those of the chondritic uniform reservoir (CHUR), an interpolation between Toluca rutile and CHUR data was used to estimate the initial $^{92}$Nb/$^{93}$Nb to $(1.6 \pm 0.3) \times 10^{-5}$. Subsequent studies (Sanloup et al., 2000; Yin et al., 2000; Münker et al., 2000) also reported $^{92}$Zr variations in meteoritic phases with fractionated Nb/Zr: rutile from the iron meteorite Zagora, zircon from the mesosiderite Chaunskij, and calcium-aluminum rich inclusions (CAIs) from the carbonaceous chondrite Allende. However, the initial $^{92}$Nb/$^{93}$Nb values inferred in the same manner were two orders of magnitude higher ($\sim 10^{-3}$). In contrast, later Zr isotopic studies of zircons from the eucrite Camel Donga (Hirata, 2001) and Allende CAIs (Schönbächler et al., 2003; Akram et al., 2013; Mane et al., 2014) did not reveal $^{92}$Zr variations due to $^{92}$Nb decay, constraining the initial $^{92}$Nb/$^{93}$Nb of the Solar System to $<10^{-4}$. Moreover, Schönbächler et al. (2002) determined internal Nb–Zr isochrons for the ordinary (H6) chondrite Estacado and the mesosiderite Vaca Muerta, and obtained initial $^{92}$Nb/$^{93}$Nb values of $\sim 10^{-5}$.

There are several possible causes of the extensive variation in the estimated initial

$^{92}$Nb/$^{93}$Nb ratios (10$^{-3}$ to 10$^{-5}$): (1) significant differences in the Nb–Zr closure age (>200 Ma) among the studied meteorites, (2) heterogeneous distribution of $^{92}$Nb in the early Solar System, (3) highly variable initial Zr isotope compositions among the studied meteorites and CHUR, and (4) analytical artifacts. To evaluate these possibilities and firmly establish the initial abundance and distribution of $^{92}$Nb in the Solar System, it is essential to define internal isochrons for multiple meteorites that originate from distinct parent bodies and whose absolute ages are precisely known. For instance, scenario (2) or (3) renders the Nb–Zr isochron regressions using CHUR and non-chondritic materials invalid. Schönbächler et al. (2002) utilized the internal isochron approach, yet the analyzed meteorites include components of different origins and their formation ages are uncertain, which prohibits a precise determination of the solar initial $^{92}$Nb abundance.

Here we present internal Nb–Zr isochrons of three unbrecciated achondrites with known U–Pb ages (Amelin et al., 2011a; Iizuka et al., 2013, 2014, 2015a): the angrite NWA 4590, the eucrite Agoult and the ungrouped achondrite Ibitira. The internal isochrons allow us to precisely determine the initial abundance of $^{92}$Nb in the Solar System and to assess its distribution in the solar nebula. We will discuss the implications of the results for early Solar System chronology and the origin of *p*-nuclei.

## 2. Samples

Northwest Africa 4590 is a very fresh angrite with plutonic igneous textures. It mainly comprises clinopyroxene, anorthite, olivine, kirschsteinite, ulvöspinel, and accessory phosphates and troilite (Kuehner and Irving, 2007). The Re–Os and highly siderophile element geochemistry indicate that while several angrites provide evidence for contamination of exogenous chondritic materials to their protoliths, NWA 4590 is the least-contaminated angrite (Riches et al., 2012). The Pb–Pb age of 4557.93 ± 0.36 Ma, which dates crystallization, was obtained from acid-leached pyroxene using the $^{238}$U/$^{235}$U value of 137.789 ± 0.021 measured for the whole rock sample (Amelin et al., 2011a). The Pb–Pb isochron age is consistent with the Mn–Cr and Hf–W isochron ages of 4557.1 ± 1.0 Ma (Yin et al., 2009) and 4557.8 ± 0.6 Ma (Kleine et al., 2009), which are calculated using the Pb–Pb isochron age of 4563.37 ± 0.25 Ma for the angrite D'Orbigny as a time anchor (all Mn–Cr and Hf–W ages in this paper are recalculated using Mn–Cr data of Glavin et al., 2004, Hf-W data of Kleine et al., 2012 and the U–Pb age of Brennecka and Wadhwa, 2012 for D'Orbigny). In addition, the phosphate Pb–Pb age (4557.381 ± 0.066 Ma; Amelin et al., 2011a) is slightly younger than that of pyroxene and constrains the cooling rate of the parent rock to 590 ± 240 K/Ma. Abundances of trace elements including Nb and Zr were previously determined by laser ablation-inductively coupled plasma mass spectrometry (LA-ICPMS) for the constitute minerals (Amelin et al., 2011b). The results demonstrate that the principal carrier phases of Zr are ulvöspinel and pyroxene, and that the former has significantly higher Nb/Zr than the latter.

Agoult is an unbrecciated, fine-grained granulitic eucrite with no chemical evidence for weathering (Yamaguchi et al., 2009). It has remnant subophitic textures composed of anhedral pyroxene and elongated plagioclase with abundant ca. 120 ˚C triple-junction, which indicates significant recrystallization of a basaltic protolith during high-grade metamorphism (Yamaguchi et al., 2009). Minor phases are opaque minerals including ilmenite and Ti-chromite, silica minerals, phosphates, troilite, zircon and baddeleyite (Yamaguchi et al., 2009). Importantly, some zircon grains in Agoult are exceptionally large (up to ~80 μm) for eucrites, allowing us to perform the first combined high-precision U–Pb and Zr isotope analyses of single meteorite zircon grains. The Agoult zircon is typically accompanied by ilmenite and tridymite, while ilmenite often includes needles of baddeleyite (Iizuka et al., 2015a). Titanium contents in these zircon grains indicate that they crystallized at subsolidus temperatures of ~900 ˚C (Iizuka et al., 2015a). The mineral assemblage and sub-solidus crystallization temperatures provide evidence that the Agoult zircon formed through $ZrO_2$ release (baddeleyite exsolution) from ilmenite followed by reaction with the surrounding silica during high-temperature metamorphism. The zircon crystallization age of 4554.5 ± 2.0 Ma was determined from eight grains with concordant U–Pb systems (Iizuka et al., 2015a), three of which were analyzed for Zr isotopes in this study. However, substantially younger dates were obtained from plagioclase and pyroxene fractions of Agoult (Iizuka et al., 2013): the plagioclase fractions gave a Pb–Pb isochron age of 4532.2 ± 1.0 Ma, whereas the pyroxene fractions yielded scattered data with Pb–Pb model ages from 4529 to 4523 Ma. Note that all of the zircon, plagioclase and pyroxene Pb–Pb ages were calculated using the whole rock $^{238}U/^{235}U$ value of 137.709 ± 0.016 (Iizuka et al., 2013). These observations indicate that Agoult underwent two metamorphic events at 4554 Ma and ca. 4530 Ma and that the latter event reset the U–Pb system in plagioclase and pyroxene, but not in zircon.

Ibitira is a fine-grained, unbrecciated vesicular basaltic achondrite. It is mainly composed of clinopyroxene (~60%) that shows coarse exsolution lamellae of augite from pigeonite and plagioclase (~30%). It also features minor tridymite, intergrowths of ilmenite and Ti-chromite, kamacite, troilite, phosphates and olivine (Wilkening and Anders, 1975; Steele and Smith, 1976). Ibitira has long been classified as an eucrite based on the mineral assemblage, but recent detailed geochemical and O isotopic studies reveal that the parent body of Ibitira is distinct from that of the eucrites (Wiechert et al., 2004; Mittlefehldt, 2005; Scott et al., 2009). The U–Pb dating of acid-leached pyroxene and whole rock fractions, together with whole rock $^{238}U/^{235}U$ measurements, yielded a Pb–Pb isochron age of 4556.75 ± 0.57 Ma, which was interpreted as the timing of the prolonged thermal metamorphism associated with the pyroxene exsolution and chemical equilibration (Iizuka et al., 2014). The Pb–Pb age is in good agreement with the Mn–Cr isochron ages of 4557.4 ± 2.5 Ma (Lugmair and Shukolyukov, 1998) and 4555.9 ± 3.2 Ma (Yin et al., 2009) defined by the whole rock, silicate minerals (essentially pyroxene) and chromite, calculated using the angrite D'Orbigny time anchor (Glavin et al., 2004; Brennecka et al., 2012). Moreover, these ages are consistent with the $^{146}Sm-^{142}Nd$ and I–Xe ages (Prinzhofer et al., 1992; Claydon et al., 2013), even though the $^{147}Sm-^{143}Nd$ systematics suggests partial Sm–Nd re-equilibration between plagioclase and phosphate (Prinzhofer et al., 1992). This likely occurred during a moderate shock event dated at 4.49 Ga by Ar-Ar chronology (Bogard and Garrison, 2003).

## 3. Methods

Three different analytical schemes were used in this study (Table 1). Analytical schemes A and B were used for fractions of pyroxene, spinel, opaque minerals, and whole rock, while scheme C was applied to zircon. The mineral fractions were hand-picked from the disaggregated specimens of NWA 4590, Agoult and Ibitira, whereas the whole rock fractions represent disaggregated materials without mineral separation. All the mineral and whole rock fractions were cleaned with distilled acetone and digested with a concentrated HF-HNO$_3$ mixture. The spinel and opaque mineral fractions were digested using a 125 ml Parr® bomb. The digested fractions were converted to a soluble form by repeated evaporation with concentrated HNO$_3$, and then dissolved in 6 M HCl. Each digestion was divided into two aliquots for the determination of the Zr isotope composition (ca. 90%) and $^{93}$Nb/$^{90}$Zr ratio (ca. 10%). The aliquots for Zr isotopic ratio measurements were processed through an ion-exchange procedure for Zr purification, whereas no chemical separation was applied to the aliquots for the $^{93}$Nb/$^{90}$Zr analyses in order to avoid chemical Nb/Zr fractionation. In analytical scheme A, Zr was separated from the matrix elements using a two-step ion-exchange procedure based on Rosholt et al. (1966) and Münker et al. (2001). The samples dissolved in 6 M HCl were loaded onto the column packed with anion exchange resin (Eichrom AG1-X8 with 200–400 mesh), in which Zr is eluted, whereas Fe and U are retained by the resin. The Zr elution was dried down, re-dissolved in 3 M HNO$_3$ + 1% H$_2$O$_2$, and loaded onto a column packed with Eichrom Ln-Spec resin (100–150 μm). Subsequently, Ti was released from the column in 3 M HNO$_3$ + 1% H$_2$O$_2$, followed by the elution of the matrix elements in 3 M and 6 M HCl as well as 0.5 M HNO$_3$. After the removal of the co-existing elements, Zr was collected in 0.5 M HNO$_3$ + 0.06 M HF. The ion-exchange chemistry in analytical scheme B consists of three-steps, the first of which is identical to that in analytical scheme A. In the second step, Zr and Hf were separated from Ti using Bio-Rad AG1-X8 (200–400 mesh) using the protocol of Schönbächler et al. (2004), in which Ti is eluted with 0.25 M H$_2$SO$_4$ + 1% H$_2$O$_2$ followed by Zr and Hf in 6 M HCl + 1 M HF. In the third step, Zr was further purified from the residual matrix including Hf using Eichrom Ln-Spec resin (100- −150 μm) following Münker et al. (2001).

Analytical scheme C was applied to the Agoult zircon. Single zircon grains were directly extracted from thin sections after back-scattered electron and cathodoluminescence imaging and ion microprobe analysis for trace elements (Iizuka et al., 2015a). The extracted grains were rinsed with 2 M $HNO_3$ and digested with a concentrated $HF$-$HNO_3$ mixture in a 125 ml Parr® bomb, followed by evaporation and dissolution in 6 M HCl. The solutions were taken to dryness and re-dissolved in 2.5 M HCl. The aliquots were processed through a two-step ion-exchange chemistry. The first step using Eichrom AG1-X8 (200–400 mesh) was designed for U–Pb dating (Iizuka et al., 2015a). Here the matrix elements including Zr were eluted in 2.5 M HCl followed by elution of Pb and U in 0.5 M $HNO_3$. The second step utilizes Eichrom Ln-Spec resin (100–150 μm) and is designed for high-precision Hf (Iizuka et al., 2015b) and Zr isotope analyses. In this step, Zr was extracted in 6 M HCl + 0.06 M HF followed by Hf in 2 M HF (Münker et al., 2001). The total procedural Zr blanks of all analytical schemes were less than 400 pg, whereas analyzed Zr fractions were more than 100 ng.

The Zr isotopic ratio measurements were performed with a Nu Plasma multi-collector-inductively coupled plasma mass spectrometer (MC-ICPMS) at The University of Manchester and a Neptune *plus* MC-ICPMS at the ETH Zürich and The University of Tokyo (Table 1). The purified Zr fractions were dissolved in 0.5 M $HNO_3$ with a trace amount of HF and introduced using a Cetac Aridus II micro-concentric desolvating nebulizer. All five Zr isotopes and $^{95}Mo$ were monitored simultaneously using an array of Faraday cups. In addition, $^{99}Ru$ as well as $^{101}Ru$ was monitored during analyses using the Neptune *plus* (analytical schemes B and C). Instrumental mass fractionation was corrected relative to $^{94}Zr/^{90}Zr = 0.3381$ (Minster and Allègre, 1982), and the interferences of $^{92,94}Mo$ on $^{92,94}Zr$ and $^{96}Mo$ and $^{96}Ru$ on $^{96}Zr$ were corrected following the protocol of Schönbächler et al. (2004). Individual sample measurements were bracketed by analyses of a synthetic Zr Alfa Aesar (#63–061671G or #33-480327B) or NIST SRM 3169 standard solution. The Zr isotopic ratios of all samples were determined relative to the standard runs and are expressed as $\varepsilon^{91,92,96}Zr$, which are the deviations of the $^{91,92,96}Zr/^{90}Zr$ ratios from the standard solution in parts per ten thousand. To evaluate the external reproducibility of the Zr isotope analyses, a terrestrial shale (Cody Shale), a basalt (BHVO-2), pyroxene, ilmenite and zircon grains (Temora-2) were analyzed under the same analytical conditions as the meteorite samples. The terrestrial sample analyses yielded a 2 s.d. of 0.21–0.49 for $\varepsilon^{91}$Zr, 0.14–0.26 for $\varepsilon^{92}$Zr, and 0.40–1.53 for $\varepsilon^{96}$Zr (Table 2).

The $^{93}$Nb/$^{90}$Zr measurements were carried out with an Elan DRC quadrupole ICPMS at the ETH Zürich. The dynamic reaction cell technique with hydrogen as a reactive gas was utilized to minimize the formation of metal-argide polyatomic ions from the sample matrix, in particular ArCr$^+$ from Cr-rich samples that interferes with Nb and Zr ions (Hattendorf et al., 2001). The analytical uncertainties on $^{93}$Nb/$^{90}$Zr shown in Table 2 reflect the external reproducibility (2 s.d.) deduced from repeated measurements of the samples.

## 4. Results

All Zr isotopic data for the meteorite and terrestrial samples are presented in the Supplementary Table and summarized in Table 2 together with the $^{93}$Nb/$^{90}$Zr ratios. In Table 2, the uncertainties of $\varepsilon^{91,92,96}$Zr are given as standard errors of the mean multiplied by Student's t factor for a 95% confidence interval (2σ). For samples that were analyzed for Zr isotopes only once, the uncertainties represent those of multiple analyses of terrestrial samples: the 2σ obtained from repeated analyses of terrestrial ilmenite for NWA 4590 spinel fractions #1 and #2, the 2σ of the Cody Shale for the Ibitira whole rock fraction, and the 2σ of the Temora-2 zircon analyses for the Agoult zircon grains AG3-Zrn#09 and AG2-Zrn#07. The $^{93}$Nb/$^{90}$Zr ratios in the analyzed fractions range from 0.0080 to 2.3 for NWA 4590, from 0.010 to 1.6 for Agoult, and from 0.012 to 1.2 for Ibitira (Fig. 1). The whole rock and silicate mineral fractions, including the three zircon grains, of all studied meteorites yield similar $\varepsilon^{92}$Zr values, ranging between -0.2 and 0.1. These values overlap with those measured for terrestrial samples. In contrast, the spinel and opaque mineral fractions with high $^{93}$Nb/$^{90}$Zr show resolved $^{92}$Zr excesses up to 0.9ε. No correlation is observed between measured $\varepsilon^{92}$Zr and $^{95}$Mo/$^{90}$Zr (Fig. 2a), indicating that the $^{92}$Zr excesses are not artifacts of the $^{92}$Mo interference but due to the radioactive decay of $^{92}$Nb. Furthermore, $\varepsilon^{92}$Zr and 1/Zr are not correlated (Fig. 2b), indicating that the regression lines in Fig. 1 are not mixing lines but isochrons. The slopes of the isochrons correspond to the initial $^{92}$Nb/$^{93}$Nb ratio at the time of Nb-Zr isotopic closure in each sample. They yield initial $^{92}$Nb/$^{93}$Nb values of (1.4 ± 0.5) × $10^{-5}$ for NWA 4590, $(8.1 \pm 3.2) \times 10^{-6}$ for Agoult, and $(1.9 \pm 1.2) \times 10^{-5}$ for Ibitira, respectively. Moreover, the y-intercepts of these isochrons define consistent initial $\varepsilon^{92}$Zr values of -0.07 ± 0.22 for NWA 4590, -0.03 ± 0.10 for Agoult and -0.17 ± 0.20 for Ibitira.

**5. Discussion**

*5.1. The initial abundance and distribution of $^{92}$Nb in the Solar System*

Previously proposed initial $^{92}$Nb/$^{93}$Nb values of the Solar System range from ~$10^{-5}$ to ~$10^{-3}$ (Harper, 1996; Sanloup et al., 2000; Münker et al., 2000; Yin et al., 2000; Hirata, 2001; Schönbächler et al., 2002). All but one of the previous estimates are based on Nb–Zr isotope data for single phases with fractionated Nb/Zr in meteorites such as zircon and rutile, combined with those for the CHUR. To evaluate the homogeneity of the initial $^{92}$Nb abundance, however, it is crucial to define internal isochrons for meteorites with known absolute ages. Using the internal isochron approach, Schönbächler et al. (2002) determined the $^{92}$Nb/$^{93}$Nb ratio at the time of Nb–Zr closure for the H chondrite Estacado and the mesosiderite Vaca Muerta. These results set an upper limit of <3 × $10^{-5}$ on the initial $^{92}$Nb/$^{93}$Nb of the Solar System. Yet a more precise estimate was hampered by the fact that these meteorites comprise components of different origins and experienced a prolonged thermal history as evidenced by the highly variable ages obtained by different decay schemes (Ireland and Woltzka, 1992; Stewart et al., 1994; Schönbächler et al., 2002; Trieloff et al., 2003; Wadhwa et al., 2003; Blinova et al., 2007; Kleine et al., 2008).

In contrast, the three unbrecciated achondrites analyzed in this study are dated with the U–Pb system and their thermal history is reasonably well constrained (section 2). Importantly, the pyroxene Pb-Pb ages of NWA 4590 and Ibitira are in excellent agreement with the internal Mn–Cr ages obtained from spinel/opaque minerals and silicate minerals (Lugmair and Shukolyukov, 1998; Yin et al., 2009). Considering that diffusion of high field strength elements including Nb and Zr is as slow as that of Cr in oxide and silicate minerals (Cherniak et al., 2007; Cherniak and Dimanov, 2010; Stenhouse et al., 2010), the Pb–Pb and Mn–Cr ages should reflect the timing of Nb–Zr closure. Thus, the present Nb–Zr isotopic data, together with the Pb–Pb ages, can be used to define the initial abundance and distribution of $^{92}$Nb in the Solar System.

For the calculation of the initial $^{92}$Nb/$^{93}$Nb of the Solar System and Nb–Zr ages, we used the $^{92}$Nb half-life of 37 ± 5 Ma (Holden, 1990). This value is based on the results of decay-counting experiments by Makino and Honda (1977) and Nethaway et al. (1978), in which $^{92}$Nb was produced by neutron irradiation of $^{92}$Mo and $^{93}$Nb, respectively. Although Audi et al. (2003) recommended a more precise value of 34.7 ± 2.4 Ma on the basis of the same experimental results, we chose the above value because it was re-calculated by applying corrections to the cross section for the neutron irradiation. Note that the choice of a $^{92}$Nb half-life does not alter the arguments below.

Combining the initial $^{92}$Nb/$^{93}$Nb value of (1.4 ± 0.5) × 10$^{-5}$ for the angrite NWA 4590 with the time interval between the CAI Pb–Pb age (4567.3 ± 0.3 Ma, Amelin et al., 2010; Connelly et al., 2012) and the NWA 4590 pyroxene Pb–Pb age (4557.93 ± 0.36 Ma, Amelin et al., 2011a), we derive an initial solar $^{92}$Nb/$^{93}$Nb value of (1.7 ± 0.6) × 10$^{-5}$ (Fig. 3). Note that the use of the slightly younger phosphate Pb–Pb age (4557.381 ± 0.066 Ma; Amelin et al., 2011a), instead of the pyroxene age, would result in an insignificant 0.02 × 10$^{-5}$ change for the initial value. Similarly, using Ibitira's initial $^{92}$Nb/$^{93}$Nb value of (1.9 ± 1.2) × 10$^{-5}$ combined with the pyroxene Pb–Pb age (4556.75 ± 0.67 Ma, Iizuka et al., 2014), an initial solar value of (2.4 ± 1.5) × 10$^{-5}$ can be derived. The eucrite Agoult with an initial $^{92}$Nb/$^{93}$Nb value of (8.1 ± 3.2) × 10$^{-6}$ and a Pb–Pb age of 4554.5 ± 2.0 Ma for zircon (Iizuka et al., 2015a) and 4532.2 ± 1.0 Ma for plagioclase (Iizuka et al., 2013) yields initial solar $^{92}$Nb/$^{93}$Nb values of (1.0 ± 0.4) × 10$^{-5}$ and (1.6 ± 0.6) × 10$^{-5}$, the latter being in better agreement with the estimates from NWA 4590 and Ibitira. These results indicate that $^{92}$Nb was homogeneously distributed among the source regions of angrites, eucrites, and the ungrouped achondrite Ibitira.

Among the three meteorites, we propose that NWA 4590 serves as the best time anchor for the Nb–Zr chronometer for the following reasons. First, NWA 4590 retains the original igneous texture and isotope signatures (Kuehner and Irving, 2007; Riches et al., 2012), while Agoult and Ibitira were subjected to chemical equilibration during high-temperature metamorphism and isotope disturbance in less robust decay systems (e.g., K–Ar) during a later thermal event (Prinzhofer et al., 1992; Bogard and Garrison, 2003; Yamaguchi et al., 2009; Iizuka et al., 2013; 2014; 2015a). Second, both the Nb–Zr and Pb–Pb ages (Amelin et al., 2011a) are most precisely known for NWA 4590. Last but not least, the discrete Pb–Pb ages of pyroxene and phosphate of NWA 4590 constrains its cooling rate to 540 ± 290 K/Ma (Amelin et al., 2011a), indicating that discrete closure temperatures for the U–Pb and Nb–Zr systems would have a negligible effect on the timescale anchoring.

Using the initial $^{92}Nb/^{93}Nb$ value and pyroxene Pb–Pb age of NWA 4590, the initial $^{92}Nb/^{93}Nb$ ratios for Estacado ($(1.2 ± 0.6) × 10^{-5}$) and Vaca Muerta ($(0.6 ± 0.3) × 10^{-5}$) (Schönbächler et al., 2002) correspond to absolute ages of 4550 ± 32 and 4514 ± 32 Ma, respectively (Fig. 4). The Nb–Zr age of Estacado is consistent with the chondrule Pb–Pb isochron age (4546 ± 18 Ma, Blinova et al., 2007) and the Hf–W isochron age (4558 ± 2 Ma, Kleine et al., 2008), but is distinctly older than the phosphate Pb–Pb age (4492 ± 15 Ma, Blinova et al., 2007) and Ar–Ar age (4435 ± 5 Ma, Trieloff et al., 2003). The Nb–Zr age of Vaca Muerta is indistinguishable from the ages based on the Mn–Cr (≤ ca. 4540 Ma, Wadhwa et al., 2003) and Sm–Nd (4530–4500 Ma; Stewart et al., 1994) decay systems, but it is distinctly younger than the zircon Pb–Pb age (4563 ± 15 Ma; Ireland and Wlotzka, 1992) and the internal Hf–W isochron age (4564.3 ± 1.4 Ma; Schönbächler et al., 2002) and older than the plagioclase K–Ar age (3.86 ± 0.07 Ga; Bajo and Nagao, 2011). These observations can be interpreted to reflect that Nb–Zr closure of the analyzed mineral phases (silicate and opaque minerals) occurred at a lower temperature than U–Pb closure in zircon, but at a higher temperature than U–Pb closure in phosphates and K–Ar closure in feldspars. This supports our usage of the NWA 4590 pyroxene Pb–Pb age rather than that of phosphate, despite the higher precision of the latter, as an anchor for the Nb–Zr chronometer.

The results of internal Nb–Zr isochron dating (Schönbächler et al., 2002; this study) are difficult to reconcile with some of the previously reported Nb-Zr isotopic data. All five internal Nb–Zr isochrons yield identical initial $\varepsilon^{92}Zr$ values (i.e., y-intercepts) at -0.1 ± 0.1 (2

s.d.), providing evidence for a homogeneous distribution of Zr isotopes among the source regions of angrites, eucrites, mesosiderites, H chondrites and the ungrouped achondrite Ibitira. Moreover, their initial $^{92}$Nb/$^{93}$Nb ratios and Pb–Pb ages are internally consistent with ~$10^{-5}$ for the initial Solar System $^{92}$Nb/$^{93}$Nb ratio (Fig. 3). Thus, Nb–Zr isochron regressions using the CHUR and single phases with fractionated Nb/Zr in those meteorites are expected to yield geologically meaningful model ages. Nevertheless, some of these regressions yielded markedly higher initial $^{92}$Nb/$^{93}$Nb values up to ~$10^{-3}$ and apparent initial $\varepsilon^{92}$Zr values as low as -4 (zircon from the mesosiderite Chaunskij and rutile from the iron meteorite Zagora (Yin et al., 2000, as well as CAIs from the chondrite Allende, Münker et al., 2000; Sanloup et al., 2000). Moreover, later Zr isotopic studies on Allende CAIs (Schönbächler et al. 2003; Akram et al 2013; Mane et al., 2014) did not identify significant $^{92}$Zr/$^{90}$Zr variations. The observed discrepancies in both the initial $^{92}$Nb/$^{93}$Nb and $^{92}$Zr/$^{90}$Zr ratios may reflect Nb and Zr isotopic heterogeneity at a mineral scale in the meteorite source regions or, alternatively, analytical artifacts.

*5.2. Potential for the early Solar System chronology*

We now explore the potential of the Nb–Zr decay system for early Solar System chronology. Given that $^{92}$Nb was homogeneously distributed in the Solar System, the Nb-Zr chronometer can be used to accurately and precisely determine crystallization and/or metamorphic ages of meteorites that contain phases with variable Nb/Zr ratios. Moreover, the Zr isotope compositions of planetary reservoirs referenced to the CHUR reflects the timescales and mechanisms of early differentiation of the planets, because both Nb and Zr are highly refractory and the initial Zr isotope composition is homogeneous in the Solar System (section 5.1). Figure 5 illustrates the extent of Zr isotopic variations that can be produced by early Solar System events, using the newly established initial Solar System abundance of $^{92}$Nb. The Zr isotope composition of a material/reservoir $j$ can be expressed as the deviation from the CHUR:

$$\Delta^{92}Zr_j = ([^{92}Zr/^{90}Zr]_j/[^{92}Zr/^{90}Zr]_{CHUR} - 1) \times 10^4 = Q_{Zr} \times [^{92}Nb/^{93}Nb]_{t0} \times f_j \times e^{-\lambda \Delta t} \qquad (1)$$

where $[^{92}Nb/^{93}Nb]_{t0} = 1.7 \times 10^{-5}$, $\lambda = 0.019$ Ma$^{-1}$ is the decay constant of $^{92}$Nb (Holden,

1990), $f_j$ = [Nb/Zr]$_j$/[Nb/Zr]$_{CHUR}$ - 1 with [Nb/Zr]$_{CHUR}$ = 0.076 (Münker et al., 2003; Schönbächler et al., 2003), and $Q_{Zr}$ = $10^4 \times$ [$^{93}$Nb/$^{92}$Zr]$_{CHUR}$ = 4.4 × $10^3$.

The model curves (Fig. 5) indicate that measurable $^{92}$Zr excesses can be preserved in rutile, if they formed within the first 100 Ma of the Solar System. In fact, rutile in the iron meteorite Toluca displays a high $\varepsilon^{92}$Zr value of 9.1 ± 1.7 with a Nb/Zr ratio of 10.0 ± 0.5 (Harper, 1996) and this yields a model age 4 ± 12 Ma after CAI formation. Notably, troilite in IAB iron meteorites exhibits extremely high Nb/Zr and, therefore, is expected to display well-resolved $^{92}$Zr excesses, even if it crystallized 200 Ma after the Solar System origin. Since rutile and troilite often co-exist within iron and stony-iron meteorites (El Goresy, 1971), these minerals can be used for internal Nb–Zr isochron dating of these meteorites. This is of significant importance because so far internal isochron dating has been carried out on few iron meteorites only (c.f., Pb–Pb dating, Blichert-Toft et al., 2010; Pd–Ag dating, Chen and Wasserburg, 1990; Horan et al., 2012; Theis et al., 2013), even though Hf–W model ages exist for many iron meteorites (e.g., Lee and Halliday, 1996; Kleine et al., 2005; Kruijer et al., 2014). These model ages, however, reflect the timing of metal–silicate segregation on the parent bodies and not the crystallization and cooling history.

In contrast to minerals with elevated Nb/Zr ratios, phases/reservoirs characterized by Nb/Zr ≤ 0.2 are not expected to display resolvable $^{92}$Zr anomalies at the current level of analytical precision. Most whole rock samples and major terrestrial and lunar silicate reservoirs posssess Nb/Zr ratios within this range and, indeed, no resolvable $^{92}$Zr anomalies are reported for terrestrial and lunar samples (e.g., Hirata, 2000; Münker et al., 2000; Schönbächler et al., 2002, 2004, 2005; Akram et al., 2015). Nevertheless, future improvements in the analytical precision may enable us to detect a Zr isotopic vestige of early planetary differentiation, particularly if a reservoir with a substantially elevated Nb/Zr (>0.2) was formed during the differentiation. On the modern Earth, specific ocean island basalts indeed feature such elevated Nb/Zr ratios, which are attributed to intra-mantle processes such as carbonate–silicate melt immiscibility or mantle fertilization by a small-degree melt (Kamber and Collerson, 2000).

*5.3. On the origin of light p-nuclei*

Our newly defined initial $^{92}$Nb abundance of the Solar System can be used to constrain the origin of $^{92}$Nb and other *p*-nuclei in the same mass range. Since $^{93}$Nb is produced by the slow (85%) and rapid (15%) neutron capture processes (Arlandini et al., 1999), normalizing $^{92}$Nb to the neighbor *p*-nucleus $^{92}$Mo is more suitable for examining the site(s) of nucleosynthesis of *p*-nuclei (Harper, 1996). Using the solar Mo/Nb ratio of 3.65 (Anders and Grevesse, 1989) yields an initial $^{92}$Nb/$^{92}$Mo ratio of 3.1 × 10$^{-5}$. Our Solar System formed in a molecular cloud that had been partially isolated from fresh nucleosynthetic inputs (e.g., Lugaro et al., 2014) and, therefore, the initial solar $^{92}$Nb/$^{92}$Mo ratio ($R_{SI}$) must be lower than the ratio in the interstellar medium ($R_{ISM}$) because of $^{92}$Nb decay. In a simplified model, the isolation effect is accounted for by adopting a free-decay interval, in which nucleosynthetic inputs cease at a certain point in time and the radionuclides decay. More plausibly, however, inputs of synthesized materials into the molecular cloud would have partially continued through mixing processes. This can be accounted for by considering following scenario (Clayton, 1983): The radionuclides were ejected from supernovae into small hot H-I (neutral hydrogen) clouds, which evaporated through the supernova shock and subsequently cooled to exchange matter with larger H-I clouds. These clouds in turn exchanged matter with molecular clouds. In this scenario, the fraction of surviving radionuclides in the early Solar System depends on the time scale for the mass exchanges ($T_{ex}$). By adopting the same parameters as those of Clayton (1983), $R_{SI}$ and $R_{ISM}$ can be related through (Rauscher et al., 2013):

$$R_{ISM} = R_{SI}[1+1.5\lambda T_{ex}+0.4(\lambda T_{ex})^2] \qquad (2)$$

In addition, since both $^{92}$Nb and $^{92}$Mo are secondary species that are synthesized from pre-existing element seed nuclei in stars, the production ratio of $^{92}$Nb to $^{92}$Mo ($R_{Prod}$) is given by:

$$R_{Prod} = \lambda T_G R_{ISM}/(k+2) \qquad (3)$$

where $T_G$ is the time between the formation of the Galaxy and our Solar System (ca. 7.5 Ga) and *k* is an integer characterizing the growth of the Galaxy: *k* = 0 corresponds to the closed-box model in which the mass of the galactic disk remains constant, whereas $k > 0$ represents open-box models involving the growth of the galactic disk by infall of metal-poor gas over longer time for larger $k$ value (Huss et al., 2009). The $k$ value of 1 or 2 has been favored by astronomical observations such as the stellar age–metallicity distribution (Clayton et al., 1988; Madau et al., 1998; Huss et al., 2009). In addition, the initial solar abundances of extinct radionuclides $^{53}$Mn, $^{107}$Pd, $^{129}$I, $^{146}$Sm, $^{182}$Hf and $^{244}$Pu can be well explained if $T_{ex}$ is within the range of 10–300 Ma (Huss et al., 2009). Thus, we take $k$ values of 1 and 2 and $T_{ex}$ values of 10, 50 and 300 Ma for the calculation of $R_{Prod}$ (Table 3). The results indicate that the production ratio of $^{92}$Nb/$^{92}$Mo is between 0.0015 and 0.035.

The long-favored mechanism for synthesis of many *p*-nuclei is photodisintegration reactions of heavier nuclei (*γ*-process) in core collapse Type II supernovae, because it reproduces well the abundances of the *p*-nuclei in the Solar System (e.g., Woosley and Howard, 1978; Rayet et al., 1995). Our estimated $^{92}$Nb/$^{92}$Mo production ratio is consistent with the production ratios predicted for the *γ*-process in Type II supernovae ($^{92}$Nb/$^{92}$Mo of 0.001–0.03, Woosley and Howard, 1978; Rayet et al., 1995; Dauphas et al., 2003) as well as Type Ia supernovae ($^{92}$Nb/$^{92}$Mo of 0.001–0.01, Howard et al., 1991; Howard and Meyer, 1993; Travaglio et al., 2014). Yet it has long been recognized that the *γ*-process in Type II supernovae significantly underproduces the *p*-nuclei of Sr, Mo, Ru, Pd and Cd due to the lack of seed nuclei in the mass region (e.g., Rayet et al., 1995). Accordingly, several different synthetic mechanisms for the light *p*-nuclei have been proposed. One possible mechanism is the α-rich freezeout in Type II supernovae (Hoffman et al., 1996): during the core collapse of a massive star, heated nuclei are broken down into nucleons and α-particles and subsequently reassemble due to rapid expansion and cooling, resulting in the synthesis of new nuclei. The α-rich freezeout can lead to the preferential synthesis of lighter *p*-nuclei, but it is predicted to yield a $^{92}$Nb/$^{92}$Mo of ~1.0 (Hoffman et al., 1996; Meyer, 2003). This ratio is significantly higher than the probable range of the production $^{92}$Nb/$^{92}$Mo ratio (Table 3).

Alternatively, the missing abundances of the light *p*-nuclei may be accounted for by the rapid proton capture (*rp*-) process in proton-rich hot matter on the surface of a neutron star (Schatz et al., 2001) or by the neutrino-induced neutron capture ($\nu p$-) process in the innermost ejected layers of Type II supernovae (Fröhlich et al., 2006; Wanajo, 2006). Note, however, that $^{92}$Nb cannot be synthesized by these processes, because it is shielded by the stable nuclide $^{92}$Mo (Dauphas et al., 2003). Thus, if either the *rp*- or *νp*-process significantly contributes to the synthesis of the light *p*-nuclei, the resultant $^{92}$Nb/$^{92}$Mo production ratio would be markedly lower than that achieved solely by the *γ*-process in Type II supernovae. Our newly defined production ratio therefore argues against extensive contribution of the *rp*- and *νp*-processes to the synthesis of the light *p*-nuclei, unless $^{92}$Nb was further selectively produced by another mechanism. Such selective $^{92}$Nb production might proceed through the neutrino process (*ν*-process), in which energetic neutrinos in Type II supernovae excite nuclei to create new nuclei through neutral- and charged-current reactions (Hayakawa et al., 2013).

More recently, it has been shown that the abundances of *p*-nuclei for both light and heavy masses can be reproduced by the *γ*-process in Type Ia supernovae, where *s*-process seeds for *p*-nuclei are readily available (Travaglio et al., 2011, 2015). Moreover, state-of-the-art calculations (Travaglio et al., 2014) predict $^{92}$Nb/$^{92}$Mo production ratios of 0.001–0.002 for the *γ*-process in Type Ia supernovae, in agreement with our estimate. The match is particularly good for a time scale of ca. 10 Ma for the mass exchanges among the hot H-I clouds, large H-I clouds, and molecular clouds (Table 3). Our results therefore support the view that the *γ*-process in Type Ia supernovae significantly contributes to the nucleosynthesis of the light *p*-nuclei (Travaglio et al., 2011, 2015), even though other processes taking place in Type II supernovae cannot be excluded.

## 6. Conclusions

The Nb-Zr isotope systematics of the angrite NWA 4590, the eucrite Agoult and the ungrouped achondrite Ibitira define initial $^{92}$Nb/$^{93}$Nb values of $(1.4 \pm 0.5) \times 10^{-5}$, $(8.1 \pm 3.2) \times 10^{-6}$ and $(1.9 \pm 1.2) \times 10^{-5}$, respectively. Considering that their Pb–Pb ages represent the timing of Nb–Zr closure, all initial values (i.e., Nb–Zr ages) are consistent with each other. This concordance suggests a homogeneous distribution of $^{92}$Nb in the early Solar System. Since NWA 4590 is the most pristine sample among the studied meteorites and its thermal history is well constrained, we recommend that its initial $^{92}$Nb/$^{93}$Nb value and pyroxene Pb–Pb age of 4557.93 ± 0.36 Ma (Amelin et al., 2011a) are used to map the Nb–Zr chronometer onto the absolute timescale. This yields a $^{92}$Nb/$^{93}$Nb ratio of (1.7 ± 0.6) × 10$^{-5}$ at the birth of our Solar System. Given this initial ratio, the Nb–Zr decay system has the potential to become a powerful tool for dating early Solar System objects containing troilite and rutile, such as iron and stony-iron meteorites. Yet, the effective application of this chronometer to early planetary differentiation will require further improvements in the precision of the Zr isotope analysis. Using the newly defined initial $^{92}$Nb abundance of the Solar System, we further estimate a nucleosynthetic $^{92}$Nb/$^{92}$Mo production ratio of between 0.0015 and 0.035, assuming that the time interval from the last nucleosynthetic input to Solar System formation is between 10 and 300 Ma. The estimated production ratio, combined with the solar abundances of the other *p*-nuclei having similar masses, suggests that the *γ*-process in Type Ia supernovae is responsible for the nucleosynthesis of the light *p*-nuclei.

**Acknowledgements**

We are grateful to M. Wadhwa and M. Zolensky for providing specimens of Ibitira. Reviews by K. Mezger and C. Münker improved the manuscript. T.I. acknowledges support from the Australian Research Council (Grant DP 1095142), the Japan Society for the Promotion of Science (Project Numbers 23840013 and 26220713) and Department of Earth and Planetary Science of the University of Tokyo (Overseas Internship Program for Outstanding Young Earth and Planetary Researchers). Y.-J.L and part of this work was supported by the European Research Council under the European Union's Seventh Framework Programme (FP7/2007–2013)/ERC Grant agreement n° [279779].

**Figure legends**

*Figure 1:* The ε$^{92}$Zr value vs. $^{93}$Nb/$^{90}$Zr ratio for whole rock and mineral fractions of (a) the angrite NWA 4590, (b) the eucrite Agoult, and (c) the ungrouped achondrite Ibitira. Error bars represent uncertainties quoted in Table 2. The isochron regressions and error envelopes (95% probability) are shown as solid lines and dashed curves, respectively. For the isochron regressions, all $^{92}$Zr/$^{90}$Zr ratios are normalized to the reference value $^{92}$Zr/$^{90}$Zr = 0.333383 for the standard solutions (Rehkämper et al., 2001). The slopes and y-intercepts of the isochrons define the initial $^{92}$Nb/$^{93}$Nb and $^{92}$Zr/$^{90}$Zr at the time of Nb-Zr closure.

*Figure 2:* (a) The ε$^{92}$Zr value vs. $^{95}$Mo/$^{90}$Zr ratio for all meteorite sample analyses. Error bars represent 2 s.e. quoted in Supplementary Table. (b) The ε$^{92}$Zr value vs. 1/Zr for all meteorite fractions analyzed in this study. Error bars represent 2σ quoted in Table 2.

*Figure 3:* The initial $^{92}$Nb/$^{93}$Nb ratio versus Pb-Pb age for five meteorites analyzed for internal Nb–Zr isochron dating (this study and Schönbächler et al., 2002). The solid line with gray bar represents the $^{92}$Nb/$^{93}$Nb decay curve starting with an initial solar $^{92}$Nb/$^{93}$Nb value of (1.7 ± 0.6) × 10$^{-5}$. The Pb–Pb age data are from Amelin et al. (2011a) for NWA 4590 pyroxene, Iizuka et al. (2014) for Ibitira pyroxene, Iizuka et al. (2015a) and Iizuka et al. (2013) for Agoult zircon and plagioclase, Blinova et al. (2007) for Estacado chondrule and phosphate, and Ireland and Wlotzka (1992) for Vaca Muerta zircon, respectively. The age of Solar System formation is adapted from the CAI Pb–Pb age (4567.3 Ma, Amelin et al., 2010; Connelly et al., 2012).

*Figure 4:* A comparison of isotopic ages for NWA 4590, Ibitira, Agoult, Vaca Muerta and Estacado. The Nb–Zr ages are calculated using the pyroxene Pb–Pb age of 4557.93 ± 0.36 Ma (Amelin et al., 2011a) and initial $^{92}$Nb/$^{93}$Nb ratio of (1.4 ± 0.5) × 10$^{-5}$ for NWA 4590. All Mn–Cr and Hf–W ages are recalculated by anchoring the $^{53}$Mn/$^{55}$Mn (Glavin et al., 2004) and $^{182}$Hf/$^{180}$Hf (Kleine et al., 2012) ratios of D'Orbigny to the Pb–Pb age (Brennecka and Wadhwa, 2012). Data sources for Nb–Zr and Pb–Pb ages are the same as in Fig. 3. The Mn–Cr age data are from Yin et al. (2009) for NWA 4590 and Ibitira, Lugmair and Shukolyukov (1998) for Ibitira and Wadhwa et al. (2003) for Vaca Muerta, respectively. The Hf–W age data are from Kleine et al. (2009) for NWA 4590, Schönbächler et al. (2002) for Vaca Muerta, and Kleine et al. (2008) for Estacado.

*Figure 5:* Variations in Zr isotope composition as function of the timing (0 to 200 Ma after Solar System formation) and degree of Nb/Zr fractionation, expressed as $\Delta\varepsilon^{92}$Zr versus Nb/Zr following Münker et al. (2000) using the newly established initial Solar System $^{92}$Nb abundance. The shaded area indicates the typical external reproducibility of Zr isotope analyses in this study (± 0.2ε). The horizontal bars and crosses at the bottom represent Nb/Zr ranges of meteoritic phases and estimated Nb/Zr values of major terrestrial and lunar silicate reservoirs, respectively: Px, pyroxene (0.006–0.024, Münker et al., 2000; this study); CAIs, Ca-Al rich inclusions (0.07–0.45, Münker et al., 2000; Schönbächler et al., 2003); Rut, rutile (3.7–13, Harper, 1996; Yin et al., 2000); Tro, troilite (>34, Humayun and Campbell, 2003); DM, depleted mantle (0.026, Salters and Stracke, 2003); MORB, mid-ocean ridge basalts (0.045, Gale et al., 2013); CC, continental crust (0.061, Rudnick and Gao, 2003); LIBC, lunar ilmenite-bearing cumulate (~0.1, Schönbächler et al., 2005).

**Table 1** Summary of the analytical schemes used in this study.

| | Scheme A | Scheme B | Scheme C |
|---|---|---|---|
| *Ion-exchange chemistry* | *1st step (AG1-X8 200–400 mesh, ~2 ml)* | *1st step (AG1-X8 200–400 mesh, ~2 ml)* | *1st step (AG1-X8 200–400 mesh, ~0.05 ml)* |
| | (1) Load sample & collect Zr (6 M HCl, ~1.5 ml) | (1) Load sample & collect Zr (6 M HCl, ~1.5 ml) | (1) Load sample & collect Zr (2.5 M HCl, 20 drops) |
| | (2) Collect Zr (6 M HCl, 4 ml) | (2) Collect Zr (6 M HCl, 4 ml) | (2) Collect Zr (2.5 M HCl, 10 drops) |
| | *2nd step (Ln-Spec 100–150 μm, ~0.5 ml)* | *2nd step (AG1-X8 200–400 mesh, ~0.7 ml)* | (3) Collect U–Pb (0.5M $HNO_3$, 23 drops) |
| | (1) Load sample (3 M $HNO_3$ + 1% $H_2O_2$, ~4 ml) | (1) Load sample (0.25 M $H_2SO_4$ + 1% $H_2O_2$, ~1.5 ml) | *2nd step (Ln-Spec 100–150 μm, ~0.1 ml)* |
| | (2) Ti removal (3 M $HNO_3$ + 1% $H_2O_2$, >7 ml) | (2) Ti elution (0.25 M $H_2SO_4$ + 1% $H_2O_2$, ~5 ml) | (1) Load sample (2.5 M HCl, ~1 ml) |
| | (3) Rinse (0.5 M $HNO_3$, 1 ml) | (3) Collect Zr & Hf (6 M HCl + 1 M HF, 3 ml) | (2) Rinse (6 M HCl + 0.06 M HF, 2 ml) |
| | (4) Rinse (3 M HCl, 1 ml) | *3rd step (Ln-Spec 100–150 μm, ~1 ml)* | (3) Collect Zr (6 M HCl + 0.06 M HF, 12 ml) |
| | (4) Rinse (6 M HCl, 3 ml) | (1) Load sample (3 M HCl, ~5 ml) | (4) Collect Hf (2 M HF, 1 ml) |
| | (3) Rinse (0.5 M $HNO_3$, 1 ml) | (2) Rinse (3 M HCl, 10 ml) | |
| | (7) Collect Zr (0.5 M $HNO_3$ + 0.06 M HF, 2.5 ml) | (3) Rinse (6 M HCl, 10 ml) | |
| | | (4) Residual Ti removal (0.5 M HCl + 1% $H_2O_2$, 20 ml) | |
| | | (5) Collect Zr (0.5 M HCl + 0.06 M HF, 8 ml) | |
| | | (6) Collect Hf (0.5 M HCl + 0.06 M HF, 10 ml) | |
| | | (7) Collect Hf (2 M HF, 1 ml) | |
| *MC-ICPMS* | | | |
| Instrument | Nu plasma (Uniersity of Manchester) | Neptune *Plus* (ETH Zürich) | Neptune *Plus* (University of Tokyo) |
| Analyte isotopes | $^{90,91}$Zr, $^{92,94}$(Zr+Mo), $^{95}$Mo, $^{96}$(Zr+Mo+Ru) | $^{90,91}$Zr, $^{92,94}$(Zr+Mo), $^{95}$Mo, $^{96}$(Zr+Mo+Ru), $^{99,101}$Ru, $^{100}$(Mo+Ru), | $^{90,91}$Zr, $^{92,94}$(Zr+Mo), $^{95}$Mo, $^{96}$(Zr+Mo+Ru), $^{99}$Ru |
| Number of cycles | 60 | 60 | 60 |
| Intergration time (s) | 5 | 4.2 | 8.4 |
| Uptake rate of nebulizer | 100 μl/min | 50 μl/min | 100 μl/min |
| Analyte Zr concentration | ~100 ppb | ~30 ppb | ~ 30 ppb |
| Bracketing standard | Zr Alfa Aesar Specpure® (#63-061671G) | SRM NIST 3169 | Zr Alfa Aesar Specpure® (#33-480327B) |

**Table 2**

Nb–Zr isotopic data for meteoritic and terrestrial samples.

| Sample | Weight (mg) | Zr (ppm) | $^{93}$Nb/$^{90}$Zr | 2 s.d. | N* | $\varepsilon^{91}$Zr | 2σ | $\varepsilon^{92}$Zr | 2σ | $\varepsilon^{96}$Zr | 2σ | Scheme[†] |
|---|---|---|---|---|---|---|---|---|---|---|---|---|
| *NWA 4590 (angrite)* | | | | | | | | | | | | |
| Whole rock fraction | 9.7 | 79 | 0.107 | 0.012 | 4 | 0.08 | 0.53 | 0.08 | 0.43 | 0.68 | 2.05 | B |
| Pyroxene fraction #1 | 4.6 | 141 | 0.011 | 0.001 | 3 | -0.21 | 0.40 | -0.12 | 0.46 | | | A |
| Pyroxene fraction #2 | 2.3 | 135 | 0.008 | 0.001 | 3 | -0.23 | 0.54 | -0.10 | 0.33 | | | A |
| Spinel fraction #1 | 0.9 | 200 | 1.870 | 0.213 | 1 | -0.19 | 0.22 | 0.57 | 0.27 | | | A |
| Spinel fraction #2 | 1.0 | 179 | 2.281 | 0.260 | 1 | -0.08 | 0.22 | 0.87 | 0.27 | | | A |
| Spinel fraction #3 | 1.0 | 118 | 1.789 | 0.204 | 3 | -0.16 | 0.95 | 0.96 | 0.37 | 1.36 | 0.87 | B |
| *Agoult (eucrite)* | | | | | | | | | | | | |
| Whole rock fraction #1 | 22.7 | 27 | 0.170 | 0.019 | 3 | 0.12 | 1.43 | -0.02 | 0.50 | | | A |
| Whole rock fraction #2 | 15.1 | 35 | 0.146 | 0.017 | 3 | 0.04 | 0.28 | -0.10 | 0.27 | | | A |
| Zircon grain AG5-Zrn#03 | | | | | 3 | 0.13 | 0.64 | -0.12 | 0.41 | 0.19 | 0.71 | C |
| Zircon grain AG3-Zrn#09 | | | | | 1 | 0.00 | 0.21 | -0.03 | 0.18 | 0.59 | 0.65 | C |
| Zircon grain AG2-Zrn#07 | | | | | 1 | 0.08 | 0.21 | 0.05 | 0.18 | 0.80 | 0.65 | C |
| Pyroxene fraction #1 | 9.3 | 14 | 0.010 | 0.001 | 3 | 0.15 | 0.15 | -0.03 | 0.37 | 0.23 | 1.24 | B |
| Opaque fraction #1 | 0.7 | 526 | 1.602 | 0.183 | 3 | -0.22 | 0.59 | 0.35 | 0.14 | | | A |
| Opaque fraction #2 | 0.8 | 417 | 1.638 | 0.187 | 5 | -0.17 | 0.07 | 0.39 | 0.16 | | | A |
| *Ibitira (ungrouped)* | | | | | | | | | | | | |
| Whole rock fraction | 5.5 | 64 | 0.134 | 0.015 | 1 | -0.09 | 0.32 | -0.23 | 0.25 | | | A |
| Pyroxene fraction #1 | 7.3 | 54 | 0.012 | 0.001 | 4 | 0.05 | 0.53 | 0.02 | 0.37 | 0.77 | 2.25 | B |
| Pyroxene fraction #2 | 4.4 | 53 | 0.012 | 0.001 | 4 | 0.07 | 0.28 | -0.03 | 0.56 | 0.56 | 1.73 | B |
| Opaque fraction #1 | 2.5 | 214 | 0.842 | 0.096 | 3 | -0.05 | 1.24 | 0.34 | 0.31 | | | A |
| Opaque fraction #2 | 2.5 | 283 | 1.176 | 0.134 | 5 | -0.14 | 0.18 | 0.50 | 0.52 | | | A |
| Opaque fraction #3 | 1.9 | 617 | 0.489 | 0.056 | 4 | -0.04 | 0.78 | 0.15 | 0.43 | 0.71 | 1.00 | B |
| *Terrestrial* | | | | | | | | | | | | |
| Cody shale | | | | | 6 | -0.17 | 0.32 | -0.05 | 0.25 | | | A[§] |
| BHVO-2 basalt | | | | | 12 | 0.18 | 0.23 | 0.04 | 0.12 | -0.12 | 0.97 | B[§] |
| Temora-2 zircon | | | | | 8 | -0.06 | 0.21 | 0.01 | 0.18 | -0.01 | 0.65 | C |
| Pyroxene | | | | | 5 | -0.06 | 0.60 | 0.17 | 0.17 | 0.25 | 0.64 | A,B |
| Ilmenite | | | | | 6 | -0.23 | 0.22 | -0.06 | 0.27 | | | A |
| Mo-doped standard solution | | | | | 3 | 0.03 | 0.30 | -0.07 | 0.13 | 0.15 | 0.80 | |

*Number of Zr isotopic ratio measurements.

[†]Analytical schemes A, B, and C correspond to those in Table 1.

[§]These terrestrial rock samples were processed for the Zr ion-exchange chromatography of Schönbächler et al. (2004). For BHVO-2, Zr separation from Hf using Ln-Spec resin was further carried out following the protocol of Münker et al. (2001).

**Table 3**

Production ratios of $^{92}$Nb to $^{92}$Mo at different $T_{ex}$ and $k$ values.

| $^{92}$Nb/$^{92}$Mo | $T_{ex}$ = 10 Ma | $T_{ex}$ = 50 Ma | $T_{ex}$ = 300 Ma |
|---|---|---|---|
| $k = 1$ | 0.0020 | 0.0042 | 0.035 |
| $k = 2$ | 0.0015 | 0.0032 | 0.026 |

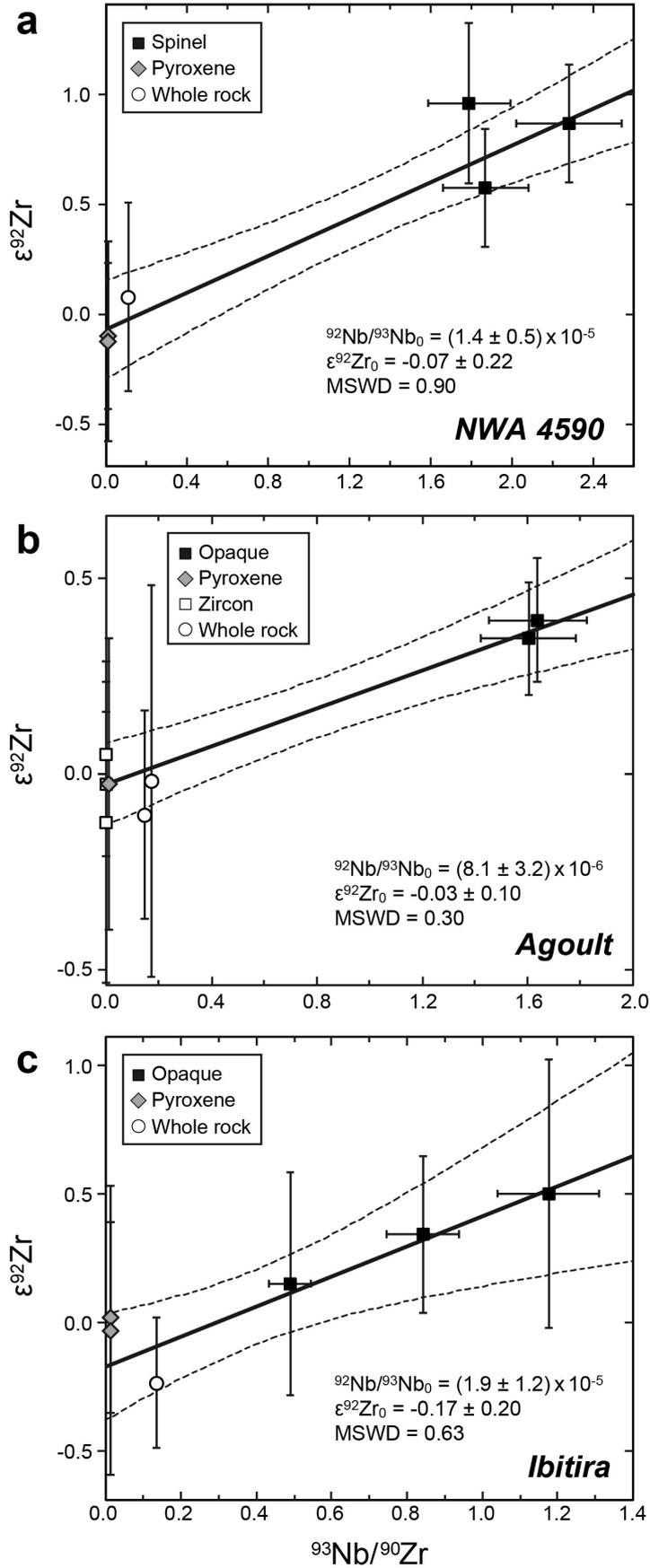

Fig. 1

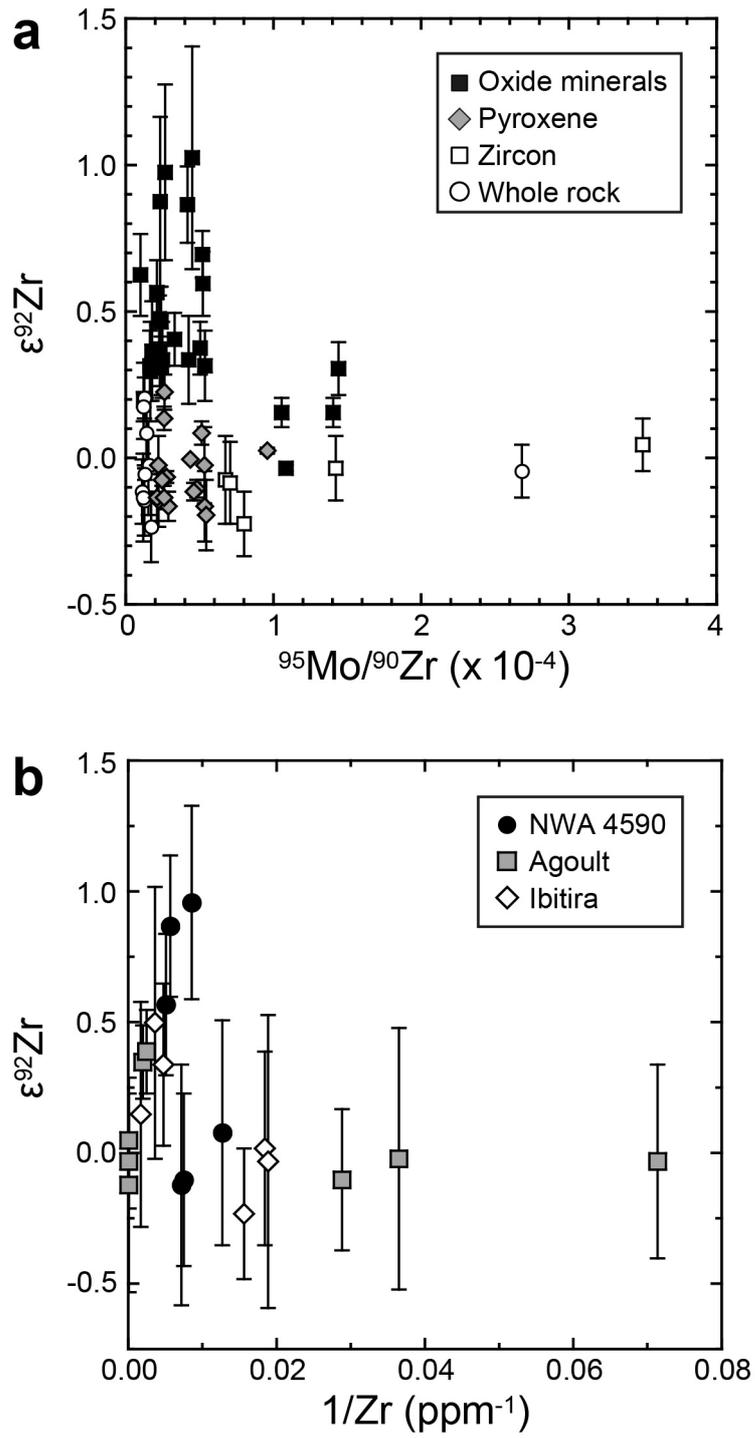

Fig. 2

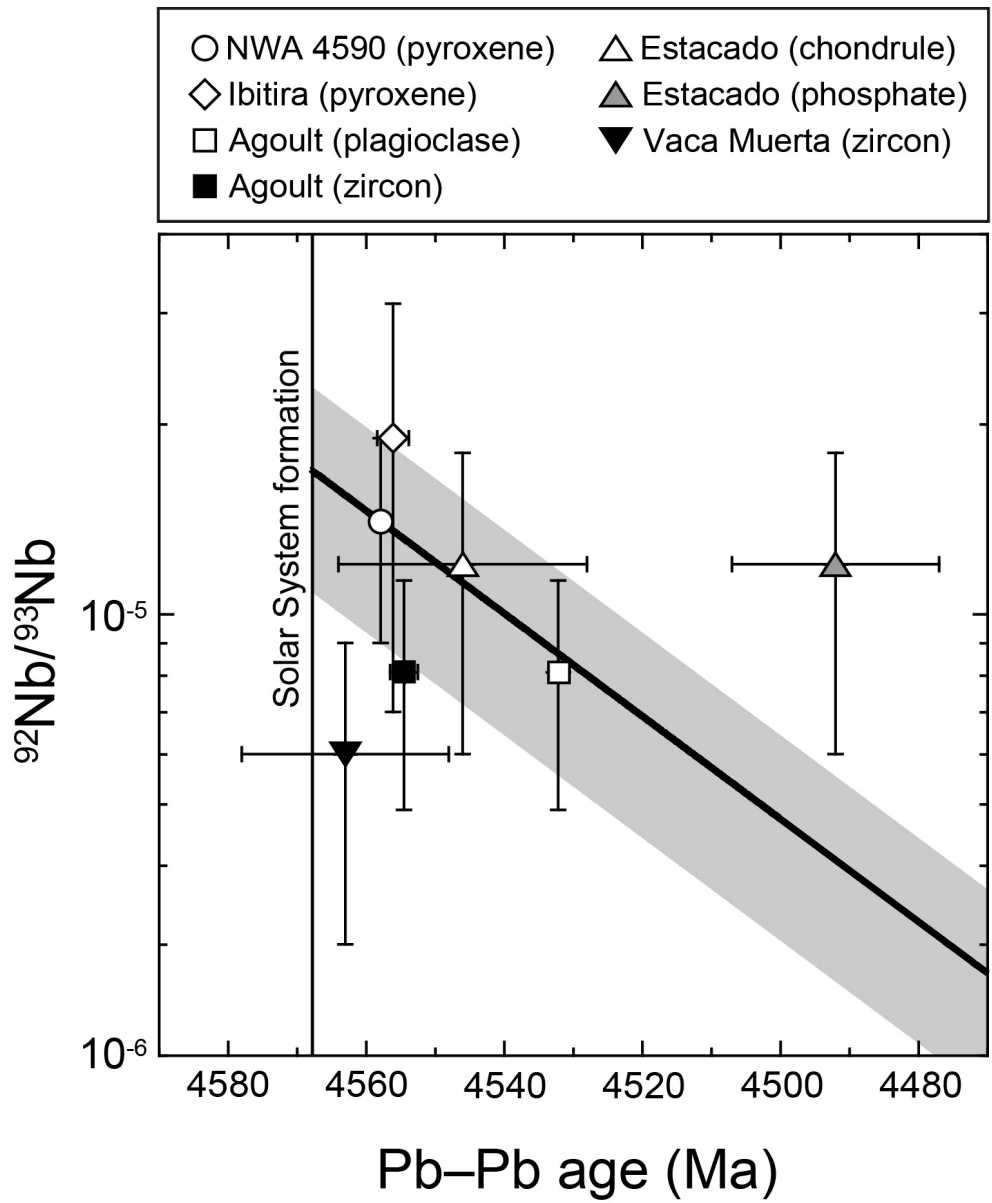

Fig. 3

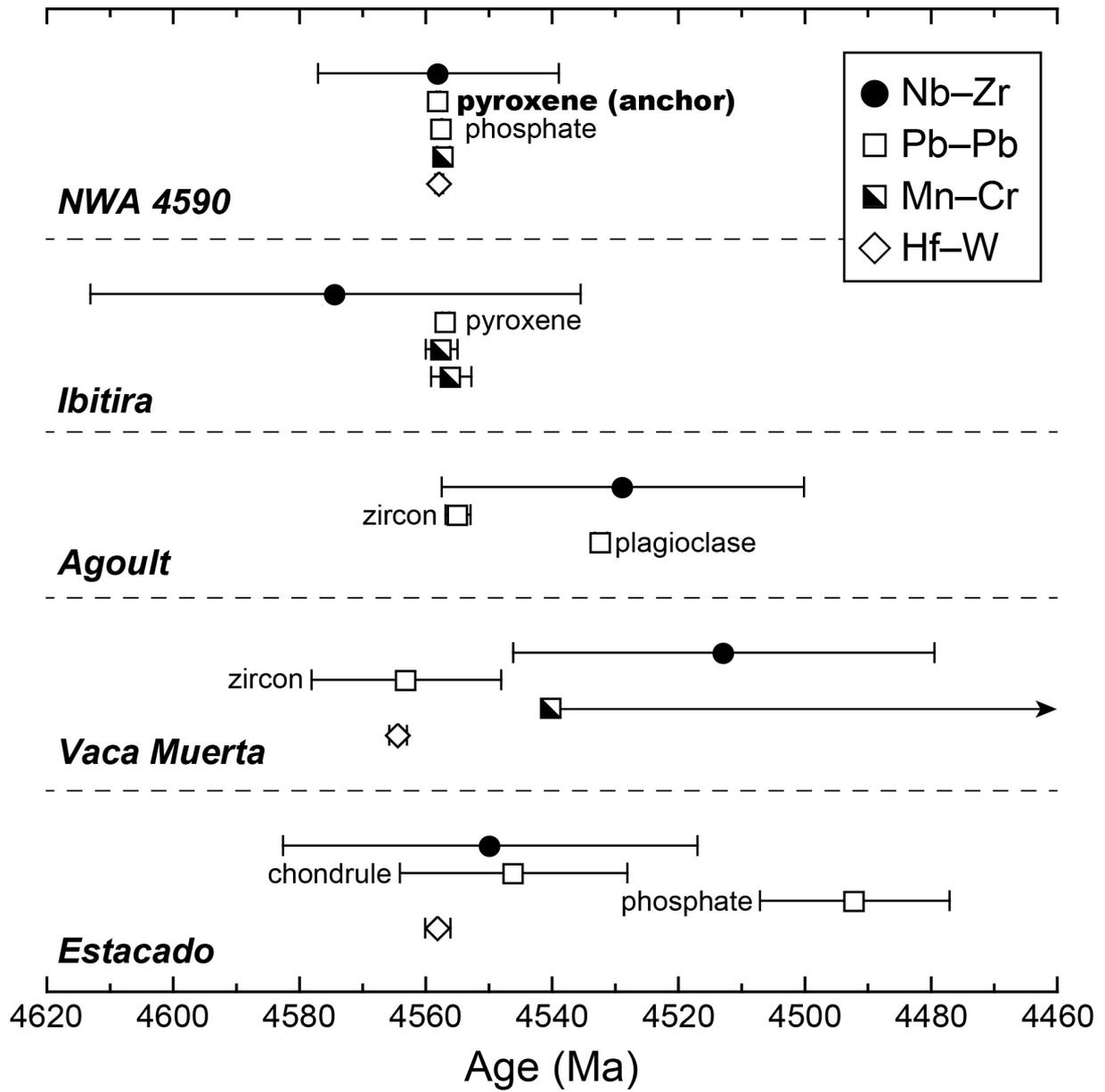

Fig. 4

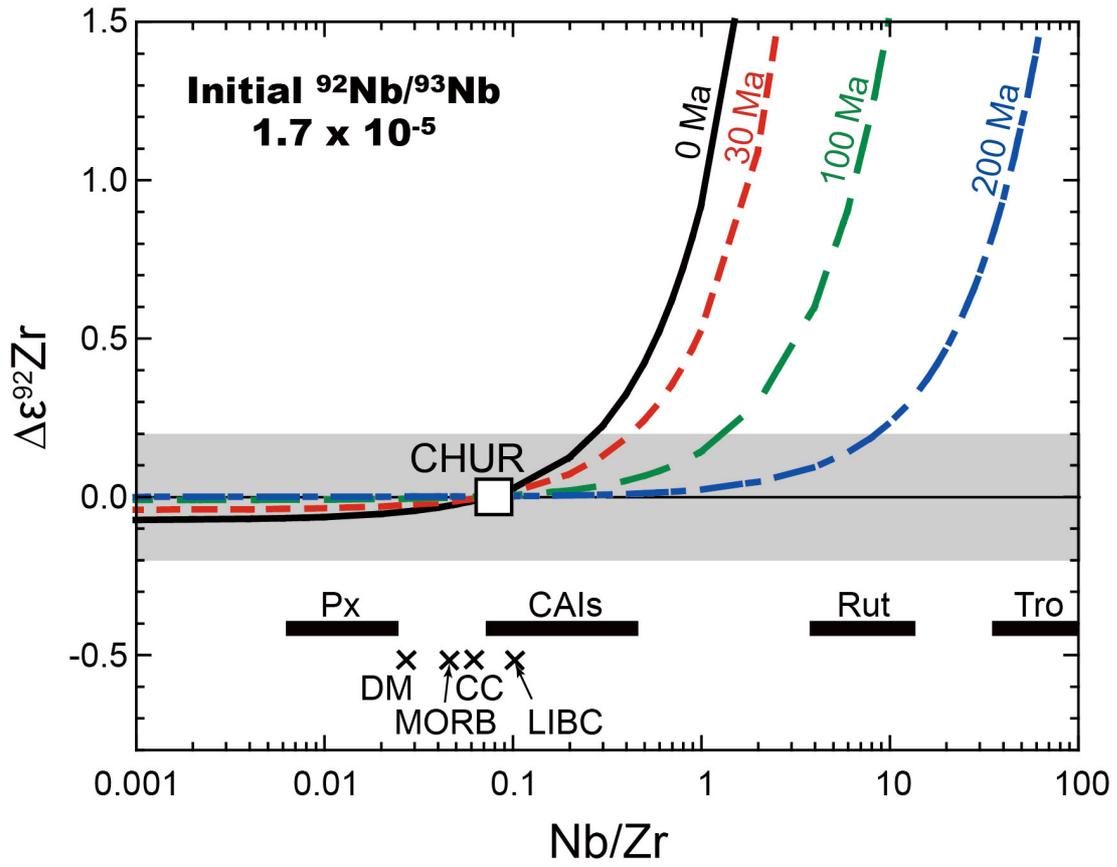

Fig. 5

**Supplementary Table**
Zr isotopic data for meteoritic and terrestrial samples.

| Sample | $\varepsilon^{96}Zr$ | 2 s.e. | $\varepsilon^{92}Zr$ | 2 s.e. | $\varepsilon^{90}Zr$ | 2 s.e. | $^{92}Mo/^{90}Zr$ | Scheme* |
|---|---|---|---|---|---|---|---|---|
| *NWA 4590 (angrite)* | | | | | | | | |
| Whole rock fraction-a | -0.05 | 0.02 | -0.02 | 0.01 | -0.10 | 0.09 | 1.5.E-05 | B |
| Whole rock fraction-b | 0.16 | 0.06 | 0.21 | 0.07 | 1.11 | 0.88 | 1.2.E-05 | B |
| Whole rock fraction-c | -0.06 | 0.02 | -0.05 | 0.02 | 1.29 | 0.95 | 1.2.E-05 | B |
| Whole rock fraction-d | 0.29 | 0.11 | 0.18 | 0.05 | 0.41 | 0.32 | 1.1.E-05 | B |
| Whole rock fraction average (± 2 s.d.) | 0.08 | 0.34 | 0.08 | 0.27 | 0.68 | 1.29 | | |
| Pyroxene fraction #1-a | -0.28 | 0.10 | -0.16 | 0.12 | | | 5.3.E-05 | A |
| Pyroxene fraction #1-b | -0.24 | 0.17 | -0.02 | 0.13 | | | 5.2.E-05 | A |
| Pyroxene fraction #1-c | -0.12 | 0.18 | -0.19 | 0.12 | | | 5.3.E-05 | A |
| Pyroxene fraction #1 average (± 2 s.d.) | -0.21 | 0.16 | -0.12 | 0.18 | | | | |
| Pyroxene fraction #2-a | -0.23 | 0.12 | -0.02 | 0.10 | | | 2.1.E-05 | A |
| Pyroxene fraction #2-b | -0.12 | 0.09 | -0.14 | 0.09 | | | 2.1.E-05 | A |
| Pyroxene fraction #2-c | -0.34 | 0.10 | -0.13 | 0.08 | | | 2.0.E-05 | A |
| Pyroxene fraction #2 average (± 2 s.d.) | -0.23 | 0.22 | -0.10 | 0.13 | | | | |
| Spinel fraction #1 | -0.19 | 0.15 | 0.57 | 0.11 | | | 2.0.E-05 | A |
| Spinel fraction #2 | -0.08 | 0.13 | 0.87 | 0.13 | | | 4.1.E-05 | A |
| Spinel fraction #3-a | 0.06 | 0.03 | 1.03 | 0.38 | 1.22 | 1.07 | 4.4.E-05 | B |
| Spinel fraction #3-b | -0.31 | 0.12 | 0.88 | 0.29 | 1.31 | 1.02 | 2.2.E-05 | B |
| Spinel fraction #3-c | -0.21 | 0.08 | 0.98 | 0.30 | 1.56 | 1.18 | 2.6.E-05 | B |
| Spinel fraction #3 average (± 2 s.d.) | -0.16 | 0.38 | 0.96 | 0.15 | 1.36 | 0.35 | | |
| *Agoult (eucrite)* | | | | | | | | |
| Whole rock fraction #1-a | 0.28 | 0.13 | 0.09 | 0.12 | | | 1.3.E-05 | A |
| Whole rock fraction #1-b | 0.31 | 0.19 | -0.03 | 0.16 | | | 1.4.E-05 | A |
| Whole rock fraction #1-c | -0.21 | 0.11 | -0.11 | 0.11 | | | 1.0.E-05 | A |
| Whole rock fraction #1 average (± 2 s.d.) | 0.12 | 0.58 | -0.02 | 0.20 | | | | |
| Whole rock fraction #2-a | 0.09 | 0.15 | -0.14 | 0.12 | | | 1.2.E-05 | A |
| Whole rock fraction #2-b | -0.02 | 0.20 | -0.13 | 0.15 | | | 1.1.E-05 | A |
| Whole rock fraction #2-c | 0.06 | 0.15 | -0.04 | 0.09 | | | 2.7.E-04 | A |
| Whole rock fraction #2 average (± 2 s.d.) | 0.04 | 0.11 | -0.10 | 0.11 | | | | |
| Zircon grain AG5-Zrn#03-a | 0.16 | 0.20 | -0.07 | 0.15 | 0.11 | 0.46 | 6.6.E-05 | C |
| Zircon grain AG5-Zrn#03-b | -0.02 | 0.19 | -0.08 | 0.14 | 0.35 | 0.37 | 7.0.E-05 | C |
| Zircon grain AG5-Zrn#03-c | 0.23 | 0.17 | -0.22 | 0.11 | 0.09 | 0.31 | 7.9.E-05 | C |
| Zircon grain AG5-Zrn#03 Average | 0.13 | 0.26 | -0.12 | 0.17 | 0.19 | 0.28 | | |
| Zircon grain AG3-Zrn#09 | 0.00 | 0.13 | -0.03 | 0.11 | 0.59 | 0.34 | 1.4.E-04 | C |
| Zircon grain AG2-Zrn#07 | 0.08 | 0.12 | 0.05 | 0.09 | 0.80 | 0.33 | 3.5.E-04 | C |
| Pyroxene fraction #1-a | 0.18 | 0.08 | 0.03 | 0.01 | -0.05 | -0.04 | 9.5.E-05 | B |
| Pyroxene fraction #1-b | 0.16 | 0.07 | -0.11 | 0.03 | 0.44 | 0.36 | 4.5.E-05 | B |
| Pyroxene fraction #1-c | 0.12 | 0.04 | 0.00 | 0.00 | 0.29 | 0.22 | 4.3.E-05 | B |
| Pyroxene fraction #1 average (± 2 s.d.) | 0.15 | 0.06 | -0.03 | 0.15 | 0.23 | 0.50 | | |
| Opaque fraction #1-a | -0.10 | 0.19 | 0.34 | 0.15 | | | 4.2.E-05 | A |
| Opaque fraction #1-b | -0.34 | 0.13 | 0.38 | 0.09 | | | 4.9.E-05 | A |
| Opaque fraction #1-c | -0.22 | 0.14 | 0.32 | 0.12 | | | 5.3.E-05 | A |
| Opaque fraction #1 average (± 2 s.d.) | -0.22 | 0.24 | 0.35 | 0.06 | | | | |
| Opaque fraction #2-a | -0.14 | 0.09 | 0.34 | 0.13 | | | 2.4.E-05 | A |
| Opaque fraction #2-b | -0.19 | 0.09 | 0.47 | 0.12 | | | 2.3.E-05 | A |
| Opaque fraction #2-c | -0.19 | 0.13 | 0.37 | 0.12 | | | 2.2.E-05 | A |
| Opaque fraction #2-d | 0.06 | 0.11 | 0.48 | 0.08 | | | 2.2.E-05 | A |
| Opaque fraction #2-e | -0.17 | 0.10 | 0.32 | 0.10 | | | 2.3.E-05 | A |
| Opaque fraction #2 average (± 2 s.d.) | -0.17 | 0.06 | 0.39 | 0.13 | | | | |
| *Ibitira (ungrouped)* | | | | | | | | |
| Whole rock fraction | -0.09 | 0.17 | -0.23 | 0.12 | | | 1.6.E-05 | A |
| Pyroxene fraction #1-a | 0.03 | 0.02 | -0.10 | 0.03 | -0.28 | -0.21 | 4.7.E-05 | B |
| Pyroxene fraction #1-b | -0.18 | 0.12 | 0.09 | 0.04 | 0.96 | 1.14 | 5.0.E-05 | B |
| Pyroxene fraction #1-c | 0.19 | 0.07 | 0.14 | 0.04 | 1.25 | 0.97 | 2.5.E-05 | B |
| Pyroxene fraction #1-d | 0.16 | 0.06 | -0.06 | 0.02 | 1.14 | 0.91 | 2.7.E-05 | B |
| Pyroxene fraction #1 average (± 2 s.d.) | 0.05 | 0.33 | 0.02 | 0.23 | 0.77 | 1.42 | | |
| Pyroxene fraction #2-a | 0.08 | 0.04 | -0.16 | 0.05 | -0.25 | -0.18 | 2.8.E-05 | B |
| Pyroxene fraction #2-b | -0.05 | 0.02 | 0.23 | 0.06 | 0.79 | 0.62 | 2.5.E-05 | B |
| Pyroxene fraction #2-c | 0.16 | 0.06 | -0.07 | 0.02 | 0.90 | 0.67 | 2.4.E-05 | B |
| Pyroxene fraction #2-d | 0.11 | 0.04 | -0.13 | 0.04 | 0.80 | 0.60 | 2.5.E-05 | B |
| Pyroxene fraction #2 average (± 2 s.d.) | 0.07 | 0.17 | -0.03 | 0.35 | 0.56 | 1.09 | | |
| Opaque fraction #1-a | -0.32 | 0.13 | 0.32 | 0.12 | | | 1.5.E-05 | A |
| Opaque fraction #1-b | 0.16 | 0.23 | 0.30 | 0.17 | | | 1.6.E-05 | A |
| Opaque fraction #1-c | 0.02 | 0.10 | 0.41 | 0.09 | | | 3.2.E-05 | A |
| Opaque fraction #1 average (± 2 s.d.) | -0.05 | 0.50 | 0.34 | 0.12 | | | | |
| Opaque fraction #2-a | -0.07 | 0.20 | 0.63 | 0.14 | | | 9.1.E-06 | A |
| Opaque fraction #2-b | -0.21 | 0.17 | 0.20 | 0.13 | | | 1.1.E-05 | A |
| Opaque fraction #2-c | -0.20 | 0.24 | 0.37 | 0.17 | | | 1.7.E-05 | A |
| Opaque fraction #2-d | -0.05 | 0.14 | 0.60 | 0.11 | | | 5.1.E-05 | A |
| Opaque fraction #2-e | -0.17 | 0.11 | 0.70 | 0.08 | | | 5.1.E-05 | A |
| Opaque fraction #2 average (± 2 s.d.) | -0.14 | 0.14 | 0.50 | 0.42 | | | | |
| Opaque fraction #3-a | 0.03 | 0.02 | 0.16 | 0.05 | 0.45 | 0.31 | 1.4.E-04 | B |
| Opaque fraction #3-b | -0.03 | 0.02 | 0.31 | 0.09 | 1.16 | 0.84 | 1.4.E-04 | B |
| Opaque fraction #3-c | 0.22 | 0.09 | 0.16 | 0.05 | 0.69 | 0.56 | 1.0.E-04 | B |
| Opaque fraction #3-d | -0.36 | 0.14 | -0.03 | 0.01 | 0.53 | 0.40 | 1.1.E-04 | B |
| Opaque fraction #3 average (± 2 s.d.) | -0.04 | 0.49 | 0.15 | 0.27 | 0.71 | 0.63 | | |
| *Terrestrial* | | | | | | | | |
| Cody shale-a | -0.08 | 0.18 | -0.14 | 0.13 | | | 8.3.E-05 | A† |
| Cody shale-b | 0.02 | 0.15 | 0.02 | 0.15 | | | 7.2.E-05 | A† |
| Cody shale-c | -0.38 | 0.17 | -0.25 | 0.12 | | | 8.0.E-05 | A† |
| Cody shale-d | -0.33 | 0.19 | -0.04 | 0.18 | | | 8.2.E-05 | A† |
| Cody shale-e | -0.11 | 0.17 | 0.05 | 0.15 | | | 6.6.E-05 | A† |
| Cody shale-f | -0.16 | 0.12 | 0.03 | 0.10 | | | 2.5.E-05 | A† |
| Cody shale average (n = 6, ± 2 s.d.) | -0.17 | 0.30 | -0.05 | 0.24 | | | 6.8.E-05 | |
| BHVO-2-a | -0.25 | 0.08 | 0.07 | 0.02 | -1.07 | 0.74 | 1.0.E-05 | B† |
| BHVO-2-b | -0.04 | 0.01 | -0.09 | 0.03 | -0.22 | 0.14 | 6.0.E-06 | B† |
| BHVO-2-c | 0.24 | 0.11 | 0.05 | 0.02 | 0.89 | 0.80 | 1.1.E-05 | B† |
| BHVO-2-d | 0.03 | 0.01 | 0.22 | 0.08 | 0.55 | 0.52 | 3.5.E-06 | B† |
| BHVO-2-e | 0.24 | 0.11 | -0.03 | 0.01 | -0.85 | 0.76 | 1.2.E-05 | B† |
| BHVO-2-f | 0.27 | 0.12 | -0.14 | 0.06 | -1.64 | 1.45 | 1.4.E-05 | B† |
| BHVO-2-g | 0.33 | 0.14 | 0.16 | 0.06 | 0.61 | 0.53 | 1.2.E-05 | B† |
| BHVO-2-h | 0.03 | 0.01 | 0.03 | 0.01 | 0.25 | 0.20 | 5.2.E-06 | B† |
| BHVO-2-i | 0.08 | 0.03 | 0.07 | 0.02 | -0.28 | 0.23 | 7.3.E-06 | B† |
| BHVO-2-j | 0.11 | 0.04 | 0.04 | 0.01 | 0.32 | 0.25 | 7.8.E-06 | B† |
| BHVO-2-k | 0.44 | 0.15 | 0.03 | 0.01 | -0.33 | 0.27 | 5.7.E-06 | B† |
| BHVO-2-l | 0.35 | 0.12 | 0.03 | 0.01 | 0.42 | 0.33 | 8.5.E-06 | B† |
| BHVO-2 basalt average (n = 12, ± 2 s.d.) | 0.18 | 0.36 | 0.04 | 0.19 | -0.12 | 1.53 | 8.6.E-06 | |
| Temora-2-a | -0.12 | 0.13 | -0.13 | 0.14 | 0.61 | 0.30 | 1.3.E-04 | C |
| Temora-2-b | -0.20 | 0.12 | -0.14 | 0.12 | 0.51 | 0.31 | 1.3.E-04 | C |
| Temora-2-c | -0.03 | 0.14 | 0.12 | 0.12 | -0.15 | 0.34 | 1.3.E-04 | C |
| Temora-2-d | -0.07 | 0.12 | 0.02 | 0.12 | -0.22 | 0.36 | 1.3.E-04 | C |
| Temora-2-e | -0.06 | 0.14 | 0.15 | 0.14 | 0.03 | 0.37 | 1.3.E-04 | C |
| Temora-2-f | 0.01 | 0.14 | 0.11 | 0.14 | -0.52 | 0.42 | 1.3.E-04 | C |
| Temora-2-g | -0.20 | 0.15 | -0.05 | 0.14 | -0.24 | 0.42 | 1.3.E-04 | C |
| Temora-2-h | 0.18 | 0.13 | 0.00 | 0.13 | -0.10 | 0.34 | 1.3.E-04 | C |
| Temora-2 zircon average (n = 8, ± 2 s.d.) | -0.06 | 0.25 | 0.01 | 0.22 | -0.01 | 0.77 | 1.3.E-04 | |
| Pyroxene-a | -0.33 | 0.10 | 0.08 | 0.11 | | | 3.7.E-05 | A |
| Pyroxene-b | -0.04 | 0.01 | 0.24 | 0.07 | 0.06 | 0.05 | 2.5.E-05 | B |
| Pyroxene-c | 0.34 | 0.11 | 0.23 | 0.07 | 0.37 | 0.26 | 2.8.E-05 | B |
| Pyroxene-d | 0.11 | 0.04 | 0.12 | 0.04 | 0.47 | 0.33 | 1.1.E-05 | B |
| Pyroxene-e | -0.28 | 0.11 | 0.18 | 0.05 | 0.08 | 0.06 | 1.1.E-05 | B |
| Pyroxene average (n = 5, ± 2 s.d.) | -0.06 | 0.49 | 0.17 | 0.14 | 0.25 | 0.40 | 2.3.E-05 | |
| Ilmenite-a | -0.22 | 0.18 | -0.12 | 0.13 | | | 1.2.E-05 | A |
| Ilmenite-b | -0.30 | 0.16 | -0.03 | 0.11 | | | 1.3.E-05 | A |
| Ilmenite-c | -0.39 | 0.15 | -0.24 | 0.14 | | | 1.2.E-05 | A |
| Ilmenite-d | -0.18 | 0.13 | 0.11 | 0.10 | | | 2.2.E-05 | A |
| Ilmenite-e | -0.10 | 0.14 | 0.05 | 0.11 | | | 2.5.E-05 | A |
| Ilmenite-f | -0.17 | 0.12 | -0.14 | 0.12 | | | 2.4.E-05 | A |
| Ilmenite average (n = 6, ± 2 s.d.) | -0.23 | 0.21 | -0.06 | 0.26 | | | 1.8.E-05 | |
| Mo-doped standard solution-a | 0.08 | 0.14 | -0.05 | 0.11 | 0.33 | 0.34 | 1.0.E-03 | |
| Mo-doped standard solution-b | -0.03 | 0.13 | -0.07 | 0.11 | 0.08 | 0.27 | 1.0.E-03 | |
| Mo-doped standard solution-c | 0.03 | 0.11 | -0.10 | 0.11 | 0.04 | 0.27 | 1.0.E-03 | |
| Mo-doped standard solution average (n = 3, ± 2 s.d.) | 0.03 | 0.12 | -0.07 | 0.05 | 0.15 | 0.32 | 1.0.E-03 | |

*Analytical schemes A, B, and C correspond to those in Table 1. These schemes yielded $^{90}Zr/^{90}Zr$ ratios of 0.333395 ± 0.000017, 0.333348 ± 0.000013, and 0.333360 ± 0.000011 for the standard solutions, respectively.

†These terrestrial rock samples were processed for the Zr ion-exchange chromatography of Schönbächler et al. (2004). For BHVO-2, Zr separation from Hf using Ln-Spec resin was further carried out following the protocol of Münker et al. (2001).